\documentclass[journal]{IEEEtran}
\usepackage{amsmath}
\usepackage{amssymb,amsfonts}
\usepackage{graphicx}
\usepackage{textcomp}
\usepackage{multicol}
\usepackage{multirow}
\usepackage{adjustbox}
\usepackage{array}
\usepackage{booktabs}
\usepackage{subfigure}
\usepackage{xcolor}
\renewcommand\arraystretch{1.08}

\newif\ifshowrev  \showrevtrue

\usepackage[maxnames=6,minnames=6,backend=bibtex,style=ieee]{biblatex}
\begin{document}

\title{Foundation Model-Driven Agentic AI for 6G}

\author{Mohammad~Ghassemi, Han~Zhang, Akram~Bin~Sediq, Ali~Afana, and~Melike~Erol-Kantarci,~\IEEEmembership{Fellow,~IEEE}%
\thanks{M.~Ghassemi, H.~Zhang, and M.~Erol-Kantarci are with the School of Electrical Engineering and Computer Science, University of Ottawa, Ottawa, ON, Canada (e-mail: \{mghas017, hzhan363, melike.erolkantarci\}@uottawa.ca).}%
\thanks{A.~Bin~Sediq and A.~Afana are with Ericsson, Ottawa, ON, Canada (e-mail: akram.bin.sediq@ericsson.com; ali.afana@ericsson.com).}\vspace{-10mm}}

\maketitle

\begin{abstract}
Mobile networks are outgrowing manual tuning: parameters multiply, services diversify, and traffic shifts faster than experts can react. The machine learning (ML) models that automate radio control are narrow: each is trained for one task, degrades when conditions move, and needs specialist retraining. This article introduces FM-in-the-loop, a framework for 6G radio access networks in which a foundation model (FM) reasons while lightweight machine learning agents act. It reads operator intents in plain language, decomposes them into subtasks, coordinates task-specific agents over Model Context Protocol (MCP) interfaces, and supervises them across their lifecycle---while the orchestration FM never enters the latency-critical control path. We locate seven integration points across four lifecycle stages, compare three cloud-to-edge deployments, and evaluate two elements: channel representations from a wireless FM are tested under noise-level and channel-estimation-error shift, and an instrumented emulation of the orchestration loop reports the measured cost of turning an operator intent into dispatched agent goals. Hallucination, cost, and standardization barriers remain.
\end{abstract}


\vspace{-1mm}
\section{Introduction}
\vspace{-1mm}

Mobile networks are planned, configured, and healed largely by human experts working through consoles and tickets. Toward 6G, the number of configurable parameters, the heterogeneity of services, and the dynamism of traffic make this operating model expensive and error-prone. Machine learning~(ML) has therefore emerged as a key enabler of autonomous networks~\cite{letaief2019roadmap}. Yet conventional ML remains narrow~\cite{xi2023rise}: a model is trained for one task on data from one deployment, degrades when conditions shift, and must be retrained by scarce specialists. Placing them in the loop helps but still depends on their availability~\cite{humanintheloop}.

\textbf{Scope.} We target the radio access network~(RAN): closed-loop optimization of beam management, mobility and handover parameters, power control, resource allocation, and reconfigurable intelligent surface~(RIS) phase configurations, at sub-second to minute timescales. Core-network functions are out of scope. We build on the open RAN~(O-RAN) architecture for a structural reason: it decomposes the RAN into vendor-neutral components---a near-real-time RAN intelligent controller~(RIC) hosting latency-critical applications, a non-real-time controller hosting slower ones, and open E2, O1, and A1 interfaces~\cite{polese2023understanding}---the plug-and-play structure an orchestrator needs to command third-party agents without vendor lock-in.

Foundation models (FMs) are pre-trained on broad datasets and adapt to many downstream tasks~\cite{bommasani2021opportunities}. Embedded in an agentic architecture~\cite{yao2022react}, an FM gives lightweight agents richer context to act on. \textbf{Terminology.} ``FM'' here covers any broadly pre-trained, widely adaptable model: language FMs, better known as large language models~(LLMs); wireless-specific FMs pre-trained on channel data; and language-action models, FMs fine-tuned so their outputs are executable tool calls rather than text. Two FM roles are kept distinct throughout: the \emph{orchestration FM}, cloud-resident and never on the latency-critical control path, and a \emph{perception FM}, a compact wireless FM embedded locally inside an agent when its measured latency satisfies the control envelope.

\textbf{Related Work and Positioning.} Four lines of work approach this problem. First, LLM-enhanced multi-agent systems coordinate several language models over networking tasks~\cite{jiang2023large}. Second, LLM-empowered edge intelligence pushes reasoning toward the edge and realizes intent-based control, in which operators state goals and the system derives the technical adjustments~\cite{shen2024large}. Third, wireless-specific FMs learn reusable channel representations from large channel corpora~\cite{alikhani2025lwm}. Fourth, and most closely related, Xiao et al.~\cite{xiao2025agentic} propose the generative FM-as-agent design, in which one FM acts directly as the network agent.

Our framework, FM-in-the-loop, is complementary to the FM-as-agent design but differs in its division of labor, as Table~\ref{tab:positioning} summarizes. Using one FM for control decisions can place its inference cost, latency, and hallucination risk on each control decision. FM-in-the-loop instead assigns reasoning to the FM and actuation to lightweight, independently verifiable agents, with the Model Context Protocol~(MCP) as the contractual boundary between them. This creates a typed, auditable FM-agent boundary for deterministic validation, keeps latency-critical control independent of orchestration-FM inference, and lets agents be retrained, replaced, or scaled independently of the FM.
The main contributions of this work are:
\begin{itemize}
\item An FM-in-the-loop framework for autonomous 6G networks that embeds FM reasoning via MCP, repositioning operator involvement toward strategic oversight.
\item A characterization of how FMs enhance ML agents at seven integration points across four lifecycle stages---requirement planning, data processing, model development, and operation (Section~III).
\item Three deployment architectures and a concrete intent schema---a structured object exchanged over MCP stating what the network should achieve, under what limits, and with what urgency (Section~II-E)---supporting cross-vendor agent invocation through a common contract.
\item A two-part evaluation: representation transfer improves RIS spectral efficiency by 24.1\% at the highest tested noise power ($-69$~dBm) over raw-CSI DDPG, and an instrumented emulation of the agentic loop reports the measured cost of turning an intent into dispatched goals.
\end{itemize}

\textbf{Why RIS?} RIS control concentrates the difficulties the framework targets: configurations grow exponentially with element count, performance is acutely sensitive to channel-estimation error and environmental change, and no hand-crafted feature set transfers across sites---a stress test for whether pre-trained representations give a lightweight agent robustness that engineered features cannot.

\begin{table}[t]
\centering
\caption{Positioning of FM-in-the-loop against related approaches to FM-driven network control.}
\label{tab:positioning}
\footnotesize
\setlength{\tabcolsep}{0.5pt}
\renewcommand{\arraystretch}{1.0}
\begin{tabular}{@{}>{\raggedright\arraybackslash}p{1.44cm}>{\raggedright\arraybackslash}p{1.52cm}>{\raggedright\arraybackslash}p{1.32cm}>{\raggedright\arraybackslash}p{1.50cm}>{\raggedright\arraybackslash}p{1.46cm}>{\raggedright\arraybackslash}p{1.43cm}@{}}
\toprule
\textbf{Approach} & \textbf{Reasoning} & \textbf{Actuation} & \textbf{FM--agent interface} & \textbf{Lifecycle governance} & \textbf{FM on critical path?} \\
\midrule
LLM-enhanced multi-agent systems~\cite{jiang2023large} & Multiple LLMs & Same LLMs & Natural-language messages & Not addressed & Yes \\
LLM-empowered edge AI and intent-based control~\cite{shen2024large} & Edge LLM & Edge LLM & Bespoke prompts & Not addressed & Yes \\
Wireless-specific FM~\cite{alikhani2025lwm} & None (perception only) & Down\-stream task model & Embedding vector & Not addressed & Perception only \\
Generative FM-as-agent~\cite{xiao2025agentic} & Generative FM & Same FM acts as agent & Internal to the FM & Emergent & Yes \\
\textbf{FM-in-the-loop (this work)} & \textbf{Cloud FM} & \textbf{ML agents} & \textbf{Typed MCP tool calls} & \textbf{Explicit, seven-point} & \textbf{No, by construction} \\
\bottomrule
\end{tabular}
\vspace{-5mm}
\end{table}

\begin{figure*}[t]
\centering
\subfigure[Human-in-the-loop framework.]
  {\includegraphics[width=3.2in]{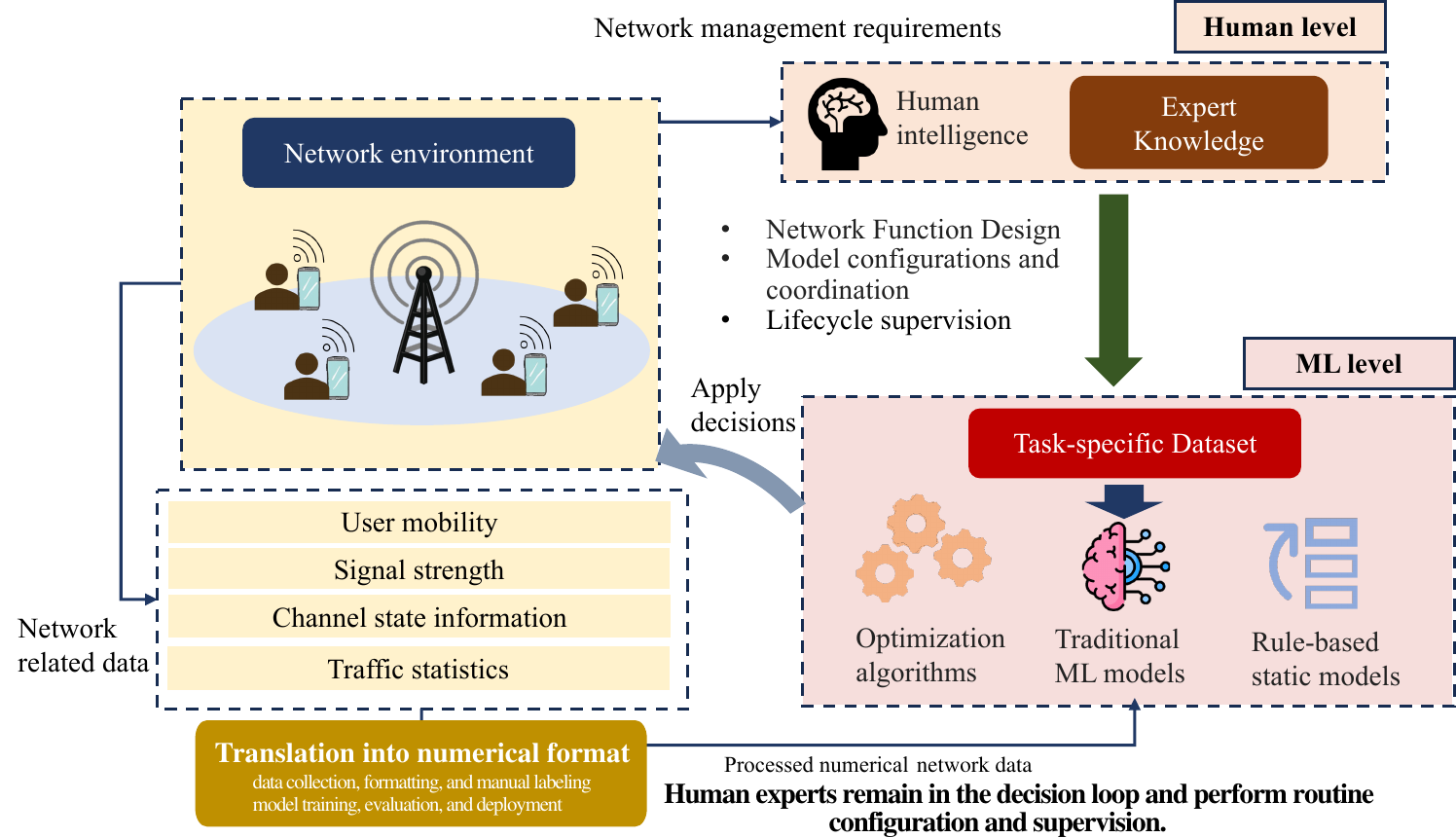}\label{fig1a}}
\hfill
\subfigure[FM-in-the-loop framework.]
  {\includegraphics[width=3.2in]{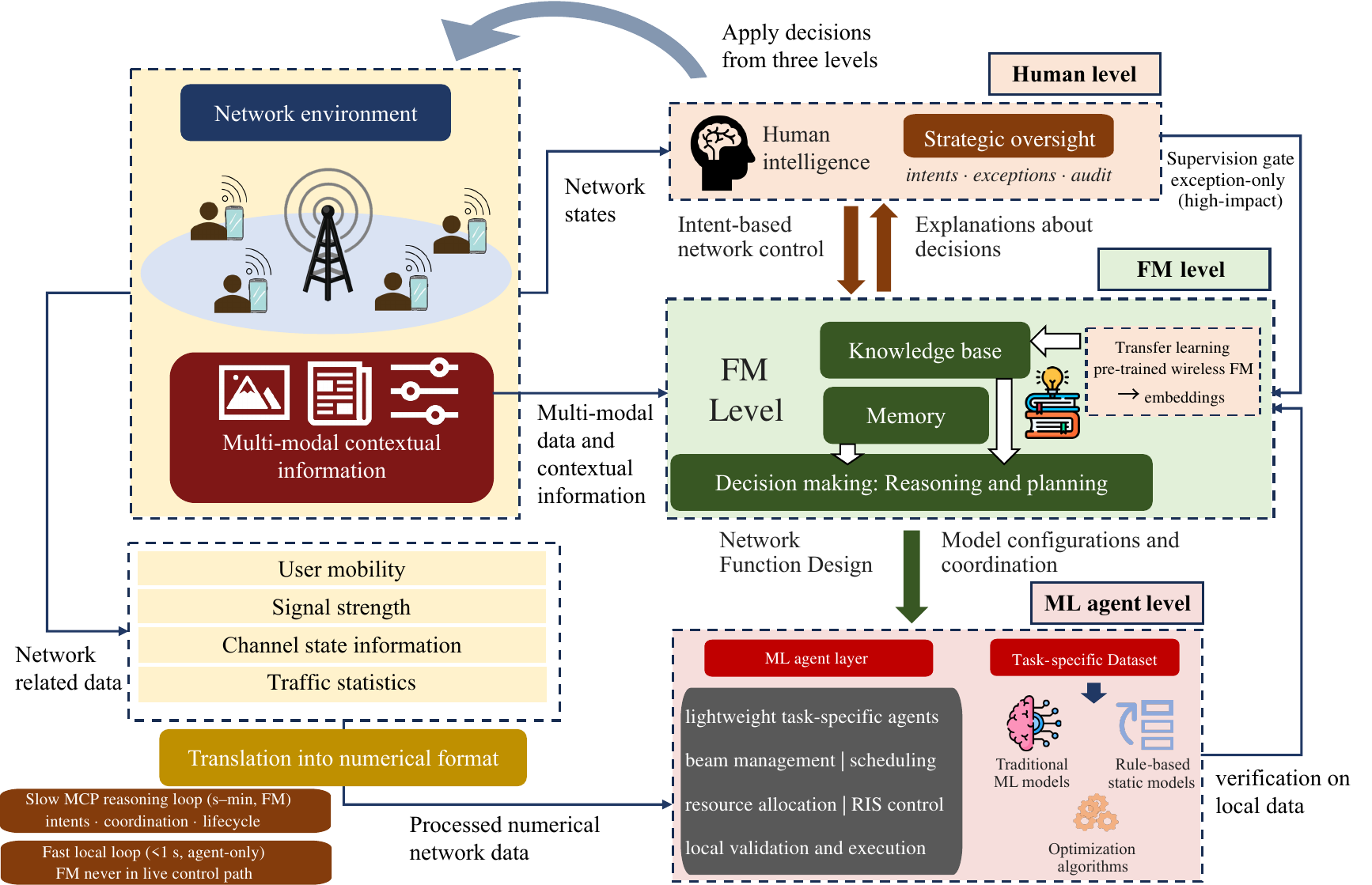}\label{fig1b}}
\caption{Evolution from human-in-the-loop to FM-in-the-loop. FMs act as intermediaries between human management and ML agents, enabling intent-driven control, human-readable reasoning traces, and self-organizing lifecycle management. In panel (b), the human role is repositioned to oversight, a single supervision gate handles exceptions only, and the two loops are drawn with their timescales; ``FM never in live control path'' refers to the orchestration FM (Section~II-A).}
\vspace{-5mm}
\label{fig1}
\end{figure*}

\vspace{-1mm}
\section{FM-in-the-Loop for Autonomous Networks}
\vspace{-1mm}

Fig.~\ref{fig1} illustrates the evolution from human-in-the-loop to FM-in-the-loop: three levels---human, FM, and ML---separate reasoning from actuation, with the FM layer realizing intent-based networking~\cite{shen2024large} under deterministic policy guardrails.

\vspace{-1mm}
\subsection{ML Agent Layer: Execution Core}
\vspace{-1mm}

The ML agent layer comprises supervised and reinforcement learning models, optimization algorithms, and rule-based strategies. It processes real-time data---user mobility, CSI, and traffic statistics---for tasks needing high update rates and deterministic outputs, exposing each agent to the FM layer through MCP-compliant interfaces for status queries, parameter updates, or retraining triggers.

\textbf{Control-loop boundary.} A hard separation between two loops is the load-bearing assumption of the design. The fast local loop---beam management, scheduling, RIS phase updates, at millisecond to sub-second timescales---runs autonomously inside edge agents without the orchestration FM; compact perception models may run locally inside an agent when latency permits (Section~I). The slow reasoning loop---intent decomposition, cross-agent coordination, lifecycle decisions, at seconds to minutes---runs in the FM layer. An orchestration FM command never actuates a radio parameter directly: it installs a policy or goal that local agents execute on their own. An orchestration FM outage therefore degrades optimization quality but does not interrupt live control.

\vspace{-1mm}
\subsection{Foundation Model Layer: Autonomous Decision Layer}
\vspace{-1mm}

The FM layer ingests multi-modal inputs including coverage maps, complaint logs, and traffic heatmaps. Two properties are central. First, transfer learning generalizes pre-trained representations to unseen conditions, reducing task-specific retraining. Second, chain-of-thought reasoning~\cite{wei2022chain} can produce intermediate text before an action is issued, but the faithfulness of that text is not assumed.

Transfer learning has two instantiations here. In cross-scenario transfer, a wireless FM pre-trained on channel data from many sites supplies representations for a new deployment, so the downstream agent trains on compact general features rather than raw high-dimensional channels; this is the path exercised in Section~V-A. In cross-task transfer, an LLM adapted on networking documents---specifications, trouble tickets, performance traces---is prompted for an unfamiliar orchestration task, avoiding task-specific model development. The rationale is the same in both: knowledge is acquired once at pre-training and amortized across deployments and tasks.

\textbf{On reasoning traces.} Chain-of-thought output is a useful audit artifact---it records what the model considered before acting, supporting root-cause analysis---but it is not a guaranteed faithful account of the model's internal computation and must not be treated as safety evidence on its own~\cite{turpin2023language}. Trust is anchored instead in verifiable mechanisms: deterministic validation of every command against specified parameter ranges, simulation-in-the-loop verification before actuation (Section~VI-C), and post-hoc consistency checks of logged traces against realized performance. Reasoning traces inform operators; they do not certify actions.

\vspace{-1mm}
\subsection{Human Layer: Strategic Oversight}
\vspace{-1mm}

Human operators supply high-level intents; the FM layer translates these into executable directives and returns human-readable explanations~\cite{humanintheloop}. The human level is thus repositioned, not removed. Rather than issuing per-cell parameter changes or labeling data by hand, the operator states intents, receives exception reports, and audits decisions after the fact, while a supervision gate retains final authority over high-impact actions such as agent replacement, by design a rare event. Manual labeling gives way to FM annotation with selective human verification (Section~III-B). The design is intended to reduce routine workload while keeping humans accountable; the present paper does not measure that reduction.

\vspace{-1mm}
\subsection{Why MCP for Multi-Vendor O-RAN?}
\vspace{-1mm}

MCP~\cite{mcp2026spec} is an open application-layer protocol---from the AI industry rather than a telecom standards body---specifying how an application discovers tools, invokes them with structured messages, and receives typed results. We adopt it as the FM--agent interface on four grounds.

\textbf{Fit to O-RAN.} MCP is a north-bound complement, not a substitute: it rides above existing control planes. In our reference mapping, MCP messages travel between the non-real-time controller, where cloud FMs reside, and agent-bearing nodes via the service exposure layer of the service management and orchestration~(SMO) framework. E2 remains the real-time radio-control path in the reference mapping, preserving the O-RAN timing model. 

\textbf{Latency.} MCP belongs exclusively to the slow reasoning loop and is tolerant of round-trip delay. We place no latency-critical function on this interface; Table~\ref{tab:overhead} therefore reports its slow-loop costs rather than a fast-loop budget.

\textbf{Security.} MCP provides protocol-level authorization mechanisms but does not confer security by design; trust must come from the carrying network and deployment controls. In the proposed deployment model, every MCP endpoint terminates inside the SMO zero-trust perimeter, with mutual transport-layer security, scoped per-invocation authorization, signed anti-replay sequence numbers, and full audit logging, aligned with O-RAN's A1/O1 security specifications~\cite{polese2023understanding}.

\textbf{Standardization status.} MCP is an openly governed application-layer protocol within the Linux Foundation's Agentic AI Foundation~(AAIF)~\cite{mcp2026spec}, but it is not part of any telecom standards baseline. O-RAN, meanwhile, has ongoing intent-driven management work for RAN services. Our proposal is therefore a research contribution that complements existing O-RAN intent-driven management rather than replacing it: the specific FM--agent contract and typed task schema proposed here can be mapped to relevant O-RAN layers depending on whether the profile concerns SMO intent management, Non-RT RIC orchestration, or Near-RT control. The multi-vendor goal is to expose agents through a common contract rather than bespoke per-pair integration.

\vspace{-1mm}
\subsection{Intent Schema and Message Flow}
\vspace{-1mm}

A minimal intent is a structured object naming the goal, the target key performance indicator~(KPI), the constraints, and the urgency:

{\footnotesize
\begin{verbatim}
{"intent_id": "INT-2041",
 "goal_type": "coverage_boost",
 "target_KPI": {"name": "cell_edge_SE",
    "operator": ">=", "value": 2.4},
 "constraint_set": {"max_tx_power_dBm": 35,
    "protected_zones": ["hospital_1"],
    "handover_success_min": 0.99},
 "priority_level": "high"}
\end{verbatim}
}

Control then crosses the MCP boundary four times with typed messages. In \textbf{validate}, an SMO-side validator checks the intent against policy rules and returns an acknowledgement or a structured rejection. In \textbf{decompose and select}, the FM queries each registered agent's capability manifest---a machine-readable list of supported tasks, input and output types, operating envelope, and recent measured performance---and selects executors. In \textbf{dispatch}, the FM issues a tool call carrying the goal as structured arguments together with timing assumptions, and the agent replies with an acceptance and its local execution plan, or rejects a cadence it cannot honor. In \textbf{monitor and audit}, progress returns as periodic notifications and every message is logged. Because timing assumptions travel inside the payload, loop-boundary violations surface at dispatch rather than at run time; Section~V-C reports the measured timings.

\begin{figure*}[t]
\centering
\includegraphics[width=6.1in]{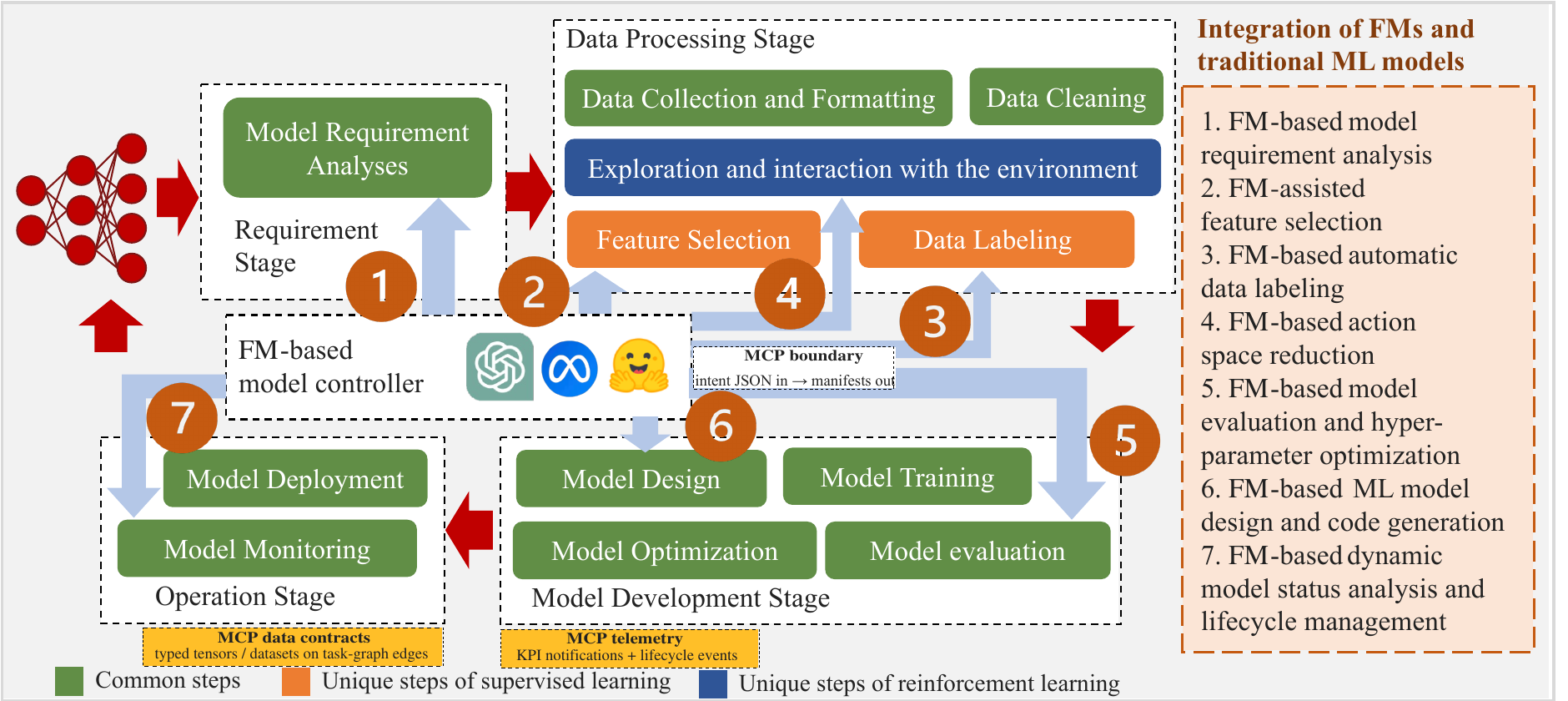}
\caption{Seven integration points between FMs and ML agents across four lifecycle stages, detailed in Section~III. Data crossing each MCP boundary is typed: intents in, capability manifests out, a task graph between Steps~1 and~2.}
\vspace{-5mm}
\label{fig2}
\end{figure*}

\vspace{-1mm}
\section{Enhancing ML Agents with Foundation Models}
\vspace{-1mm}

FMs and ML agents have fundamentally different lifecycles. FMs are pre-trained on large datasets, expensive to retrain, and adapted by lightweight fine-tuning; ML agents are lightweight, task-specific, and frequently retrained. Fig.~\ref{fig2} shows seven integration points across four stages. They form one pipeline: Step~1 turns an intent into a task graph whose nodes are candidate agents, Steps~2--3 fill each node's data requirements, Steps~4--6 build and configure each node's model, and Step~7 keeps the graph healthy in production, feeding drift back to Step~1.

\vspace{-1mm}
\subsection{Requirement Stage: Task Planning (Step~1)}
\vspace{-1mm}

Requirement analysis, traditionally performed by domain experts, is automated by FMs that decompose high-level tasks into subtasks and assign suitable agents. Decomposition proceeds in three moves: parse the intent into a structured goal graph; query each agent's MCP capability manifest to find matching executors, falling back to code generation for uncovered subtasks; and emit a task graph annotated with dependencies and data flows, which is passed to Steps~2--3 for data preparation and subsequently used by Step~7 for lifecycle coordination.

For example, the intent ``maximize cell-edge user throughput in Cluster~7 without degrading handover success below 99\%'' decomposes into load-prediction, resource-allocation, and mobility-robustness subtasks. Handover control and resource allocation are tightly coupled---an aggressive allocation change induces boundary users to churn cells---so the framework marks them a coordinated pair: they exchange action proposals over a shared MCP channel each decision round and negotiate before committing, while the FM supervises only their joint performance and arbitrates on escalation. Fast mutual adaptation thus stays inside the agent layer.

\vspace{-1mm}
\subsection{Data Processing: Feature Selection and Labeling (Steps~2--3)}
\vspace{-1mm}

FMs can assist semantic feature selection from task descriptions and network context, reducing dependence on hand-designed features, and their annotation can reduce manual labeling; selective human verification remains advisable for safety-critical tasks until reliability bounds are established.

\vspace{-1mm}
\subsection{Model Development: Design and Optimization (Steps~4--6)}
\vspace{-1mm}

FMs accelerate model design through automated code generation, which must pass formal testing before carrier-grade deployment, and interpret evaluation results to adjust architectures and hyperparameters. On detecting performance collapse under high-Doppler conditions, an FM can introduce sequence-modeling layers to address temporal non-stationarity. For action-space reduction, FMs prune beamforming candidates using channel geometry from a perception FM, or propose smaller candidate sets for RL agents with large action spaces; the resulting exploration benefit must be measured per task.

Concretely, the FM maintains per-agent configuration records covering architecture family, learning rates, replay-buffer sizing, exploration parameters, and retraining cadence. When validation performance violates targets, the FM proposes a bounded configuration patch, searching only within ranges validated for that agent class, deploys it in shadow mode against the digital twin, and promotes it only after the twin confirms improvement. Hyperparameter adjustment is therefore closed-loop and auditable rather than one-shot generation.

\vspace{-1mm}
\subsection{Operation Stage: Lifecycle Management (Step~7)}
\vspace{-1mm}

After deployment, the FM monitors model status: when concept drift---a shift in user distribution or channel statistics---degrades an agent, it triggers retraining or a registry swap, and balances performance against compute cost by deactivating intensive agents in low-traffic periods, unlike static rule-based management requiring manual reconfiguration.

Drift detection applies a configurable per-agent two-sample test to live input and output distributions against a rolling training-time reference, alarming when significant divergence is sustained over a monitoring window. The present emulation records the detector event only; production test family, threshold, and window length require a deployment study. The response ladder is graduated: recalibrate normalization statistics; trigger targeted fine-tuning on recently buffered data through low-rank adapters; if recovery fails within a bounded number of attempts, hot-swap to the best-matching registered alternative from the manifests of Step~1; and fall back to a rule-based safety policy if no substitute exists. Every rung is logged as an MCP-style event, giving operators a complete lifecycle audit trail. Section~V-C exercises this ladder.

\vspace{-1mm}
\section{Deployment Architectures}
\vspace{-1mm}

Deploying large FMs directly on O-RAN distributed units or radio units remains challenging because of thermal and memory constraints. Three practical architectures follow.

Fully centralized. Both layers reside on cloud servers, enabling coordinated decisions for load balancing, interference management, and slicing, at the cost of limited scalability in ultra-dense networks, round-trip latency, and privacy exposure when telemetry leaves the site.

Fully distributed. Distilled FMs and agents sit at edge or local sites, improving privacy and scalability since data stays local, and negotiate actions without central arbitration, aligning with the near-real-time and non-real-time controllers~\cite{polese2023understanding}---at the cost of higher local resource demand and synchronization complexity.

Hybrid: cloud reasoning, edge execution. FMs reside in the cloud while lightweight or distilled agents run on edge nodes. The cloud FM issues policy directives over MCP---beam index sets and per-user power weights for a venue under a power budget---and local agents perform millisecond-level beam steering without further cloud interaction. This is the architecture used in Section~V.

\vspace{-1mm}
\section{Case Study}
\vspace{-1mm}

We evaluate two elements separately, since they carry different kinds of evidence. Section~V-A isolates representation transfer: a wireless FM supplying robust state features to a lightweight reinforcement-learning executor, exercising the cross-scenario transfer path of Section~II-B under the hybrid deployment. Section~V-C exercises the agentic loop in an instrumented local emulation and reports its measured cost. Section~V-D states what remains unvalidated.

\vspace{-1mm}
\subsection{Representation Transfer for RIS Control}
\vspace{-1mm}

We consider a RIS-assisted downlink in an urban environment subject to blockage: a base station with 32 antennas, a RIS with 32 passive reflecting elements establishing quasi line-of-sight paths for obstructed users, and 10 single-antenna users. Beamforming and RIS phase shifts are jointly optimized to maximize spectral efficiency under a transmit power budget and a unit-modulus constraint on each element.

The Large Wireless Model~(LWM)~\cite{alikhani2025lwm}, a transformer pre-trained on large-scale wireless channel data by self-supervision, is adopted as the perception FM inside the control agent: it maps raw channel matrices to contextualized representations that encode spatial and spectral structure. It is deliberately compact---12~layers, embedding size~64, 603{,}712 parameters and 2.30\,MiB at single-precision floating point~(FP32)---and measured at low inference latency on the test platform (Table~\ref{tab:overhead}), supporting feasibility for local deployment subject to target-hardware validation. Rather than selecting actions itself, LWM supplies latent channel representations to the control agent. Because LWM was not trained on RIS-aided environments, we fine-tune it by masked self-supervision, masking 15\% of input patches and reconstructing them. The representations serve as state inputs to a deep deterministic policy gradient~(DDPG) agent whose actor-critic design fits joint phase-shift and beamforming control.

Setup. We use the ``O1'' outdoor scenario of the DeepMIMO dataset~\cite{alkhateeb2019deepmimo} (unrelated to the O-RAN O1 interface), a ray-traced urban channel model at 28\,GHz with 0.5\,GHz bandwidth, with users served at 35\,dBm. LWM is fine-tuned with AdamW at learning rate 0.00005, weight decay 0.00001, and batch size~64. The baseline agent is DDPG with two hidden layers of 256 rectified-linear units in actor and critic, learning rate 0.001, discount factor 0.99, soft target update 0.001, a replay buffer of 100{,}000 transitions, batch size 128, and Ornstein--Uhlenbeck exploration noise decaying over the first 10{,}000 steps. The raw-CSI, engineered-CSI, zero-shot, and random-projection DDPG variants use the same action space and reward and differ in state representation. The robust soft actor-critic~(SAC), beam-sweeping, transformer-policy, and optimization-based RIS baselines are evaluated on the same task and operating points but are not representation-only ablations: the transformer policy replaces the multilayer-perceptron actor with a transformer that maps state directly to actions, and the optimization-based controller recomputes phase shifts from the available channel estimate at each decision step instead of learning a policy. Each configuration runs over five random seeds for 400 episodes of 200 steps; we report means. We vary noise power from $-85$\,dBm to $-69$\,dBm and channel-estimation signal-to-noise ratio~(SNR) over 0, 5, 10, 15, and 20~dB. The comparison uses nine methods: fine-tuned LWM-aided DDPG, raw-CSI DDPG, engineered-CSI DDPG, zero-shot LWM-aided DDPG, random-projection-aided DDPG, robust SAC, beam sweeping, the transformer policy, and the optimization-based RIS controller.

\vspace{-1mm}
\subsection{Results}
\vspace{-1mm}

\begin{figure}[t]
\centering
\includegraphics[width=\columnwidth]{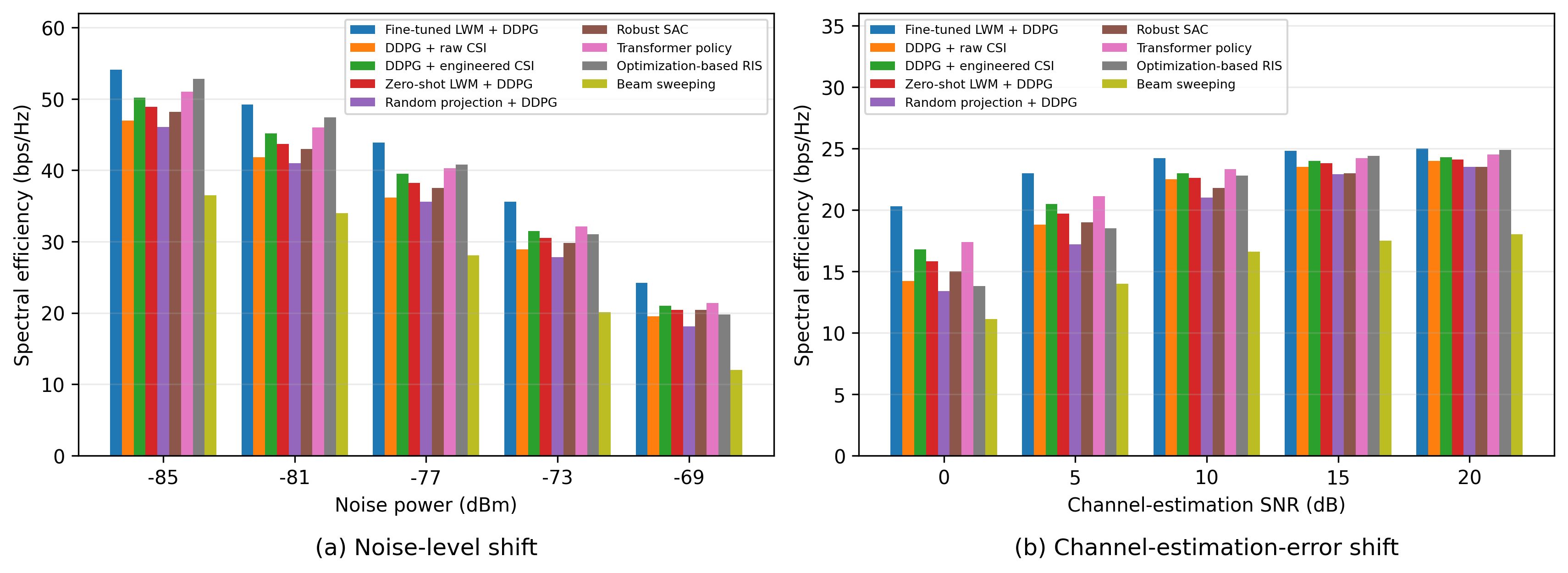}
\caption{Measured sum spectral efficiency under noise-level shift (a) and channel-estimation-error shift (b). The DDPG variants provide controlled representation comparisons; robust SAC, beam sweeping, the transformer policy, and the optimization-based RIS controller are controller-level reference baselines under the same operating points. Values are means over the five-seed evaluation.}
\vspace{-7mm}
\label{fig3}
\end{figure}

Fig.~\ref{fig3} reports the measured mean sum spectral efficiency for the nine comparison methods. In panel~(a), the fine-tuned LWM-aided DDPG reaches 24.2~bps/Hz at $-69$~dBm, compared with 19.5~bps/Hz for raw-CSI DDPG, a 24.1\% improvement. In panel~(b), it reaches 20.3~bps/Hz at 0~dB channel-estimation SNR versus 14.2~bps/Hz for raw-CSI DDPG, and the gap narrows to 4.2\% at 20~dB. The two controller-level references behave differently. The transformer policy reaches 21.4~bps/Hz at $-69$~dBm---above raw-CSI DDPG but 12\% below the fine-tuned LWM---so added policy capacity recovers only part of the representation gain. The optimization-based controller remains close to the fine-tuned LWM where CSI is accurate (52.8 vs.\ 54.1~bps/Hz at $-85$~dBm) but degrades as estimation error grows, to 13.8~bps/Hz at 0~dB channel-estimation SNR against 20.3~bps/Hz.

\textbf{What ``distribution shift'' means here.} The experiment covers noise power and channel-estimation error. In deployed networks, shift also arises from mobility and Doppler spread, blockage dynamics, topology changes, user-density fluctuation, and carrier-frequency differences. We summarize channel quality with an \emph{effective SNR}: the ratio between received signal power and everything that corrupts it---thermal noise, interference, and channel-estimation error. We use effective SNR only as a common scalar measure of observed channel quality; it does not capture the full structure of other distribution shifts, which remain future work. This is also why a pre-trained model such as LWM is used: structural shifts recur across environments and can be learned from pre-training data, whereas noise power cannot be learned away, since it bounds the information the channel carries. Robustness to noise therefore provides a deliberately challenging test, which we evaluate directly. The claim we support is correspondingly narrow: LWM representations retain their measured performance under the tested noise-level and channel-estimation-error conditions on the DeepMIMO O1 scenario.

\vspace{-1mm}
\subsection{The Agentic Loop: From Intent to Dispatch}
\vspace{-1mm}

To exercise the omitted elements, we instrument the slow-loop path in a local emulation harness. Llama-3.1-8B-Instruct receives the structured operator intent together with the two-stage filtered capability manifests and produces a structured decomposition and agent-selection result; deterministic checks validate the result before dispatch. The registry and agent endpoints are emulated, and the message exchanges follow the typed MCP-style structure described in Section~II-E rather than constituting a wire-level multi-vendor MCP interoperability test. Three framework behaviors are observed rather than asserted. First, the guardrail rejects intents whose constraints violate specified parameter ranges, and an agent rejects a dispatched goal whose decision period it cannot honor, so loop-boundary violations surface at dispatch. Second, the handover and resource-allocation agents converge their proposals through direct negotiation, without an FM round-trip per exchange. Third, injected drift triggers the two-sample detector and the response ladder of Section~III-D walks from recalibration to hot-swap, each rung logged. The emulation records the detection event under the same caveats as Section~III-D.

Table~\ref{tab:overhead} reports costs measured in this local emulation harness: fast loop on an NVIDIA RTX 3090 GPU; slow loop on Llama-3.1-8B-Instruct via vLLM on the same workstation. The fast loop costs 0.03~ms for the DDPG actor and 0.18/0.55~ms for the LWM perception pass at batch size 1/64 per control step---0.21~ms combined at batch size 1 and 0.58~ms at batch size 64. The slow loop runs per intent or lifecycle event: the 95th-percentile decomposition and selection time is 0.85~s, while manifest matching was measured at 0.09~ms for the tested 4- and 100-agent registries. Two-stage deterministic filtering---first by task family, then by published recent performance---bounds FM context growth, which otherwise grows linearly with registry size, giving 760 tokens per intent with 4 agents and 820 with 100. Transport and O-RAN interoperability overhead are excluded.

\begin{table}[t]
\centering
\caption{Operational cost by control loop (local emulation, Section~II-A).}
\label{tab:overhead}
\footnotesize
\setlength{\tabcolsep}{3pt}
\renewcommand{\arraystretch}{1.0}
\begin{tabular}{@{}>{\raggedright\arraybackslash}p{3.4cm}>{\raggedright\arraybackslash}p{1.0cm}>{\raggedright\arraybackslash}p{1.5cm}>{\raggedright\arraybackslash}p{1.85cm}@{}}
\toprule
\textbf{Stage} & \textbf{Loop} & \textbf{Basis} & \textbf{Cost} \\
\midrule
DDPG actor & Fast & per step & 0.03\,ms \\
LWM representation & Fast & batch 1/64 & 0.18 / 0.55\,ms \\
LWM weight storage, FP32/INT8/INT4 & --- & theoretical & 2.30 / 0.58 / 0.29\,MiB \\
\midrule
Intent validation & Slow & per intent & 0.02\,ms \\
Decomposition + selection & Slow & per intent & 0.85\,s \\
\quad manifest matching & Slow & per intent & 0.09\,ms \\
MCP-style dispatch & Slow & per intent & 3.4\,ms \\
Tokens per intent, 4/100 agents & Slow & per intent & 760 / 820 \\
Twin validation & Slow & per command & 85\,ms \\
\midrule
LoRA fine-tuning & Offline & per drift & 0.35\,GPU\,h \\
\bottomrule
\multicolumn{4}{@{}p{7.6cm}@{}}{\footnotesize Fast-loop entries are mean per-step times; slow-loop entries are per-intent or per-command measurements, except decomposition and selection, which is the 95th percentile over 500 intents.  Transport and O-RAN interoperability overhead excluded. GPU: graphics processing unit; FP32/INT8/INT4: 32-bit floating-point and 8-/4-bit integer quantization; LoRA: low-rank adaptation.}
\vspace{-5mm}
\end{tabular}
\end{table}

\vspace{-1mm}
\subsection{Scope of Evidence and Roadmap}
\vspace{-1mm}

Section~V-A validates representation transfer under the two tested shift axes. Section~V-C exercises planning, selection, coordination, and lifecycle management on an emulated registry, not independent production applications. Multi-vendor field interoperability and workload reduction remain unvalidated; the next step is an O-RAN-grade testbed with live traffic.

\vspace{-1mm}
\section{Challenges and Potential Solutions}
\vspace{-1mm}

\subsection{Limited Domain Knowledge}
\vspace{-1mm}

Most general-purpose FMs lack mobile-communications knowledge; wireless-specific FMs such as LWM address this at the channel level but must extend to the full management stack. Remedies include in-context learning, retrieval-augmented generation---retrieving the relevant specification when validating a beamforming command against regulatory power limits~\cite{zhang2024large}---and low-rank adaptation, which freezes pre-trained weights and injects trainable low-rank matrices at a fraction of full retraining's cost.

\vspace{-1mm}
\subsection{Computation Cost and Inference Delay}
\vspace{-1mm}

Large FMs impose significant fine-tuning and inference overheads. Mitigations include distillation into edge-deployable agents; event-triggered inference that queries the FM only when an anomaly detector flags a deviation; and hardware-aware quantization with sparse attention, for which the theoretical weight-storage footprint falls from 2.30\,MiB at FP32 to 0.58\,MiB with 8-bit integer~(INT8) weights and 0.29\,MiB with 4-bit integer~(INT4) weights, with no architectural change. Table~\ref{tab:overhead} reports the measured costs by control loop. On the fast loop, the LWM representation pass takes 0.18/0.55~ms at batch 1/64 and the DDPG actor takes 0.03~ms, giving a combined measured inference time of 0.21~ms at batch size 1 and 0.58~ms at batch size 64 per control step. On the slow loop, the 95th-percentile decomposition and selection time is 0.85~s and an emulated MCP-style dispatch round trip takes 3.4~ms, incurred once per intent rather than per control step. The hybrid deployment keeps the fast loop local and invokes the orchestration LLM only per intent, keeping FM inference off the fast control path.

\vspace{-1mm}
\subsection{Robustness and Hallucination Risks}
\vspace{-1mm}

Unlike text generation, FM hallucinations in network control---an occupied frequency, an exceeded power limit---can cause outages or violations. We propose a verify-then-act mechanism: the agent layer simulates FM-issued commands in a reduced-fidelity digital twin before applying them. Twin validation is reserved for slow-loop commands, whose decision cycles of seconds to minutes tolerate the measured 85~ms validation time (Table~\ref{tab:overhead}); latency-critical functions are never routed through it (Section~II-A). Twin fidelity is calibrated online against realized outcomes, and a command that cannot be validated within its timing envelope is rejected to a safe fallback rather than applied on trust. Deterministic validators enforce specified parameter ranges at the FM--execution interface. Adversarial robustness remains an open area.

\vspace{-1mm}
\subsection{Standardization and Interoperability}
\vspace{-1mm}

Multi-vendor scalability requires agreed interfaces. O-RAN provides established management and control interfaces~\cite{polese2023understanding}, but it does not define the specific FM--agent contract and typed task schema proposed here. Our proposal therefore complements, rather than replaces, existing O-RAN management and control mechanisms. The profile can be mapped to relevant O-RAN layers depending on whether it concerns SMO intent management, Non-RT RIC orchestration, or Near-RT control; MCP remains an application-layer research choice rather than an O-RAN baseline.

\vspace{-1mm}
\section{Conclusion}
\vspace{-1mm}
We proposed the FM-in-the-loop framework for autonomous 6G networks, embedding FM reasoning into the agent lifecycle through MCP. Two results are demonstrated, each under limited conditions. In simulation, fine-tuned wireless FM representations deliver 24.1\% higher spectral efficiency at $-69$~dBm than raw-CSI DDPG. On a single workstation, the orchestration loop completes intent validation, agent selection, agent negotiation, and drift-triggered hot-swapping at the measured costs of Table~\ref{tab:overhead}. End-to-end architectural integration, wire-level multi-vendor O-RAN interoperability, and measured operator-workload reduction remain envisioned capabilities, pending an O-RAN-grade testbed with live traffic. Three barriers define the agenda: hallucination risk calls for verify-then-act guardrails backed by verifiable evidence rather than raw reasoning traces; limited wireless domain knowledge motivates in-context learning, retrieval augmentation, and low-rank adaptation; and inference cost demands distilled agents with event-triggered activation.

\vspace{-1mm}
\section*{ACKNOWLEDGMENT}
\vspace{-1mm}
We honor the memory of our co-author, Dr. Han Zhang, whose dedication and substantive contributions shaped this article.

\begin{refcontext}[sorting=none]
\small
\printbibliography

@article{letaief2019roadmap,
  author  = {Letaief, Khaled B. and Chen, Wei and Shi, Yuanming and
             Zhang, Jun and Zhang, Ying-Jun Angela},
  title   = {The Roadmap to {6G}: {AI} Empowered Wireless Networks},
  journal = {IEEE Communications Magazine},
  volume  = {57},
  number  = {8},
  pages   = {84--90},
  year    = {2019},
}

@article{humanintheloop,
  author  = {Yan, Qi and Chen, Wei and Poor, H. Vincent},
  title   = {Big Data Driven Wireless Communications: A Human-in-the-Loop
             Pushing Technique for {5G} Systems},
  journal = {IEEE Wireless Communications},
  volume  = {25},
  number  = {1},
  pages   = {64--69},
  year    = {2018},
}

@article{bommasani2021opportunities,
  author  = {Bommasani, Rishi and Hudson, Drew A. and Adeli, Ehsan and
             Altman, Russ and Arora, Simran and von Arx, Sydney and
             Bernstein, Michael S. and others},
  title   = {On the Opportunities and Risks of Foundation Models},
  journal = {arXiv:2108.07258},
  year    = {2021},
  note    = {[Online]. Available: https://arxiv.org/abs/2108.07258.
             Accessed on: Sep.\ 7, 2026}
}

@article{xi2023rise,
  author  = {Xi, Zhiheng and Chen, Wenxiang and Guo, Xin and He, Wei and
             Ding, Yiwen and Hong, Boyang and others},
  title   = {The Rise and Potential of Large Language Model Based Agents:
             A Survey},
  journal = {Science China Information Sciences},
  volume  = {68},
  number  = {2},
  pages   = {121101},
  year    = {2025},
}

@inproceedings{yao2022react,
  author    = {Yao, Shunyu and Zhao, Jeffrey and Yu, Dian and Du, Nan and
               Shafran, Izhak and Narasimhan, Karthik and Cao, Yuan},
  title     = {{ReAct}: Synergizing Reasoning and Acting in Language Models},
  booktitle = {Proc. Int. Conf. Learning Representations (ICLR)},
  address   = {Kigali, Rwanda},
  month     = may,
  year      = {2023}
}

@article{jiang2023large,
  author  = {Jiang, Feibo and Peng, Yubo and Dong, Li and Wang, Kezhi and
             Yang, Kun and Pan, Cunhua and Niyato, Dusit and
             Dobre, Octavia A.},
  title   = {Large Language Model Enhanced Multi-Agent Systems for
             {6G} Communications},
  journal = {IEEE Wireless Communications},
  volume  = {31},
  number  = {6},
  pages   = {48--55},
  year    = {2024},
}

@inproceedings{alikhani2025lwm,
  author    = {Alikhani, Sadjad and Charan, Gouranga and Alkhateeb, Ahmed},
  title     = {{LWM}: A Pre-trained Wireless Foundation Model for
               Universal Feature Extraction},
  booktitle = {Proc. IEEE Int. Conf. Machine Learning for Communication
               and Networking (ICMLCN)},
  address   = {Barcelona, Spain},
  pages     = {1--6},
  month     = may,
  year      = {2025},
}

@article{shen2024large,
  author  = {Shen, Yifei and Shao, Jiawei and Zhang, Xinjie and
             Lin, Zehong and Pan, Hao and Li, Dongsheng and
             Zhang, Jun and Letaief, Khaled B.},
  title   = {Large Language Models Empowered Autonomous Edge {AI} for
             Connected Intelligence},
  journal = {IEEE Communications Magazine},
  volume  = {62},
  number  = {10},
  pages   = {140--146},
  year    = {2024},
}

@article{xiao2025agentic,
  author  = {Xiao, Yong and Shi, Guangming and Zhang, Ping},
  title   = {Toward Agentic {AI} Networking in {6G}: A Generative
             Foundation Model-as-Agent Approach},
  journal = {IEEE Communications Magazine},
  volume  = {63},
  number  = {9},
  pages   = {68--74},
  year    = {2025},
}

@misc{mcp2026spec,
  author       = {{Model Context Protocol Project}},
  title        = {Model Context Protocol Specification,
                  Version 2026-07-28},
  year         = {2026},
  howpublished = {[Online]. Available:
                  https://modelcontextprotocol.io/specification/2026-07-28},
  note         = {A Series of LF Projects, LLC. Accessed on: Sep.\ 7}
}

@article{wei2022chain,
  author  = {Wei, Jason and Wang, Xuezhi and Schuurmans, Dale and
             Bosma, Maarten and Ichter, Brian and Xia, Fei and Chi, Ed
             and Le, Quoc V. and Zhou, Denny},
  title   = {Chain-of-Thought Prompting Elicits Reasoning in Large
             Language Models},
  journal = {Advances in Neural Information Processing Systems},
  volume  = {35},
  pages   = {24824--24837},
  year    = {2022}
}

@article{turpin2023language,
  author  = {Turpin, Miles and Michael, Julian and Perez, Ethan and
             Bowman, Samuel R.},
  title   = {Language Models Don't Always Say What They Think:
             Unfaithful Explanations in Chain-of-Thought Prompting},
  journal = {Advances in Neural Information Processing Systems},
  volume  = {36},
  pages   = {74952--74965},
  year    = {2023}
}

@article{polese2023understanding,
  author  = {Polese, Michele and Bonati, Leonardo and D'Oro, Salvatore
             and Basagni, Stefano and Melodia, Tommaso},
  title   = {Understanding {O-RAN}: Architecture, Interfaces, Algorithms,
             Security, and Research Challenges},
  journal = {IEEE Communications Surveys \& Tutorials},
  volume  = {25},
  number  = {2},
  pages   = {1376--1411},
  year    = {2023},
}

@inproceedings{alkhateeb2019deepmimo,
  author    = {Alkhateeb, Ahmed},
  title     = {{DeepMIMO}: A Generic Deep Learning Dataset for Millimeter
               Wave and Massive {MIMO} Applications},
  booktitle = {Proc. Information Theory and Applications Workshop (ITA)},
  address   = {San Diego, CA, USA},
  month     = feb,
  year      = {2019},
  note      = {[Online]. Available: https://arxiv.org/abs/1902.06435.
               Accessed on: Sep.\ 7, 2026}
}

@inproceedings{zhang2024large,
  author    = {Zhang, Han and Bin Sediq, Akram and Afana, Ali and
               Erol-Kantarci, Melike},
  title     = {Large Language Models in Wireless Application Design:
               In-Context Learning-enhanced Automatic Network Intrusion
               Detection},
  booktitle = {Proc. IEEE Global Communications Conference (GLOBECOM)},
  address   = {Cape Town, South Africa},
  pages     = {2479--2484},
  month     = dec,
  year      = {2024},
}
\end{refcontext}

\vspace{-12mm}
\begin{IEEEbiographynophoto}{Mohammad Ghassemi}
is pursuing the Ph.D. degree at the University of Ottawa. His interests include machine learning for wireless communications, foundation models, and autonomous network management.
\end{IEEEbiographynophoto}
\vspace{-12mm}

\begin{IEEEbiographynophoto}{Han Zhang}
received the Ph.D. degree from the University of Ottawa. Her research applies large language models to wireless network management.
\end{IEEEbiographynophoto}
\vspace{-12mm}

\begin{IEEEbiographynophoto}{Akram Bin Sediq}
(Senior Member, IEEE) is a Principal AI/ML Developer at Ericsson and an Adjunct Research Professor at Carleton University.
\end{IEEEbiographynophoto}
\vspace{-12mm}

\begin{IEEEbiographynophoto}{Ali Afana}
(Senior Member, IEEE) is a 5G Systems Developer at Ericsson. His interests include wireless communications and signal processing.
\end{IEEEbiographynophoto}
\vspace{-12mm}

\begin{IEEEbiographynophoto}{Melike Erol-Kantarci}
(Fellow, IEEE) is Canada Research Chair in AI-enabled Next-Generation Wireless Networks and Full Professor at the University of Ottawa, and Strategic Product Manager for AI in RAN at Ericsson.
\end{IEEEbiographynophoto}

\end{document}